\documentclass[10pt,conference]{IEEEtran}

\usepackage{cite}
\usepackage{amsmath,amssymb,amsfonts}
\usepackage{algorithmic}
\usepackage{graphicx}
\usepackage{textcomp}
\usepackage{xcolor}

\begin{document}

\makeatletter
\newcommand{\linebreakand}{%
  \end{@IEEEauthorhalign}
  \hfill\mbox{}\par
  \mbox{}\hfill\begin{@IEEEauthorhalign}
}
\makeatother

%\title{Full Stack Quantum Computing in Practice: Ecosystems, Stakeholders and Challenges}
\title{Full-Stack Quantum Software in Practice: Ecosystem, Stakeholders and Challenges}

% Authors
\author{
\IEEEauthorblockN{Vlad Stirbu}
\IEEEauthorblockA{%\textit{Department of XYZ} \\
\textit{University of Jyväskylä}\\
Jyväskylä, Finland \\
vlad.a.stirbu@jyu.fi}
\and
\IEEEauthorblockN{Majid Haghparast}
\IEEEauthorblockA{%\textit{Department of XYZ} \\
\textit{University of Jyväskylä}\\
Jyväskylä, Finland \\
majid.m.haghparast@jyu.fi}
\linebreakand
\IEEEauthorblockN{Muhammad Waseem}
\IEEEauthorblockA{%\textit{Department of XYZ} \\
\textit{University of Jyväskylä}\\
Jyväskylä, Finland \\
muhammad.m.waseem@jyu.fi}
\and
\IEEEauthorblockN{Niraj Dayama}
\IEEEauthorblockA{%\textit{Department of XYZ} \\
\textit{University of Jyväskylä}\\
Jyväskylä, Finland \\
niraj.r.dayama@jyu.fi}
\and
\IEEEauthorblockN{Tommi Mikkonen}
\IEEEauthorblockA{%\textit{Department of XYZ} \\
\textit{University of Jyväskylä}\\
Jyväskylä, Finland \\
tommi.j.mikkonen@jyu.fi}
% \and 
% \IEEEauthorblockN{Author 3}
% \IEEEauthorblockA{\textit{Department of XYZ} \\
% \textit{University of ABC}\\
% City, Country \\
% email1@example.com}

% \and
% \IEEEauthorblockN{Author 4}
% \IEEEauthorblockA{\textit{Department of XYZ} \\
% \textit{University of ABC}\\
% City, Country \\
% email2@example.com}
}

\maketitle

% Abstract
\begin{abstract}
The emergence of quantum computing has introduced a revolutionary paradigm capable of transforming numerous scientific and industrial sectors. Nevertheless, realizing the practical utilization of quantum software in real-world applications presents significant challenges. Factors such as variations in hardware implementations, the intricacy of quantum algorithms, the integration of quantum and traditional software, and the absence of standardized software and communication interfaces hinder the development of a skilled workforce in this domain. This paper explores tangible approaches to establishing quantum computing software development process and addresses the concerns of various stakeholders. By addressing these challenges, we aim to pave the way for the effective utilization of quantum computing in diverse fields.
\end{abstract}

% Keywords

\maketitle

\section{Introduction}

Quantum computing holds great promise as a revolutionary technology that has the potential to transform various fields. By harnessing the principles of quantum mechanics, quantum computers can perform complex calculations and solve problems that are currently intractable for classical computers. This promises breakthroughs in areas such as cryptography, optimization, drug discovery, materials science, and machine learning. Quantum computing's ability to leverage quantum mechanics properties like superposition, interference and entanglement can unlock exponential speedups and enable more accurate simulations of quantum systems.

The development of quantum software faces numerous challenges that need to be addressed for harnessing the power of quantum computing effectively. Firstly, the limited availability and instability of quantum hardware pose significant obstacles. Quantum computers are prone to errors and noise, necessitating the development of robust error correction techniques. Additionally, quantum programming languages and tools are still in their nascent stages, requiring advancements to facilitate efficient software development. Furthermore, the scarcity of skilled quantum software developers and a lack of standardization hinder the widespread adoption of quantum software. As quantum systems scale, the complexity of designing and optimizing quantum algorithms increases, demanding novel approaches to algorithm design and optimization. Addressing these challenges is crucial for realizing the full potential of quantum computing and enabling the development of practical quantum software applications.

%\waseem{The summary of the related work should  go here.....}

This paper explores the challenges and approaches to establishing a quantum computing software development process. It highlights the obstacles in realizing practical utilization of quantum software, such as hardware variations, algorithm complexity, integration with traditional software, and the lack of standardized interfaces. Furthermore, the paper emphasizes the need to address these challenges to enable effective utilization of quantum computing.

\section{Background}
\label{background}

\subsection{Qubit implementation}

The current candidates for building general-purpose quantum computers, as listed in Table \ref{tab:quantum_qubit_tech}, fall under the category of Noisy Intermediate-Scale Quantum (NISQ) systems. Although these quantum computers are not yet advanced enough to achieve fault-tolerance or reach the scale required for quantum supremacy, they provide an experimentation platform to develop new generations of hardware, develop quantum algorithms and validate quantum technology in real world usecases. Whether a quantum computer is general-purpose or specialized, the selection of quantum qubit implementation technology can significantly enhance hardware efficiency for specific problem classes. To make effective use of the hardware, application developers must consider these differences when designing and optimizing the software's functionality and operations.

\begin{table*}[htbp]
\caption{Quantum Computing Qubit Implementation Technologies}
\label{tab:quantum_qubit_tech}
\centering
\begin{tabular}{|l|p{7cm}|p{7cm}|}
\hline
\textbf{Qubit Technology} & \textbf{Description} & \textbf{Applicability} \\
\hline
Superconducting & Tiny superconducting materials are cooled to extremely low temperatures to manifest their quantum properties. & General-purpose quantum computing, suitable for various types of problems. \\
\hline
Trapped Ion & Ions are trapped within electromagnetic fields. & General-purpose quantum computing, with potential for high coherence and low error rates. \\
\hline
Topological & A new approach to quantum computing that leverages the properties of topological states of matter to create qubits. Unlike other qubit technologies, which typically rely on individual particles like ions or electrons, topological qubits are based on collective properties of an ensemble of particles. & General-purpose quantum computing, aimed at achieving fault-tolerant operations. \\
\hline
Photonic & Quantum information is stored in photons that can be manipulated and transmitted over long distances. & General-purpose quantum computing, suitable for communication and cryptography applications. \\
\hline
Annealing & Special purpose quantum computers designed to solve optimization problems. & Specialized quantum computing, specifically targeted at optimization and sampling problems. \\
\hline
\end{tabular}
\end{table*}

\subsection{Quantum algorithms}

%[Outline of algorithms that promise advantages over classical approaches. Not too deep, just references. The intent is to introduce algorithm developers as special kind of stakeholders, different than software developers.]

Quantum algorithms are computational techniques specifically designed to harness the unique properties of quantum systems \cite{Montanaro2016}. They offer significant advantages over classical algorithms in certain computational tasks. One key advantage is the ability to solve complex problems exponentially faster. For example, Shor's algorithm enables efficient factoring of large numbers, posing a potential threat to current encryption methods. Also, Grover's algorithm provides substantial speedup in searching large databases. Moreover, quantum algorithms can address optimization problems more effectively, leading to improved solutions in areas like portfolio optimization, logistics, and drug discovery.

\subsection{Software}

A typical quantum program performs a specialized task as part of a larger classical program, see Fig. \ref{fig:model}. The quantum program is submitted as a batch task to a classical computer that controls the operation of the quantum computer. The classical computer schedules the task execution and provides the result to the classical program when the job completes.

\begin{figure}
  \centering
  \includegraphics[width=0.45\textwidth]{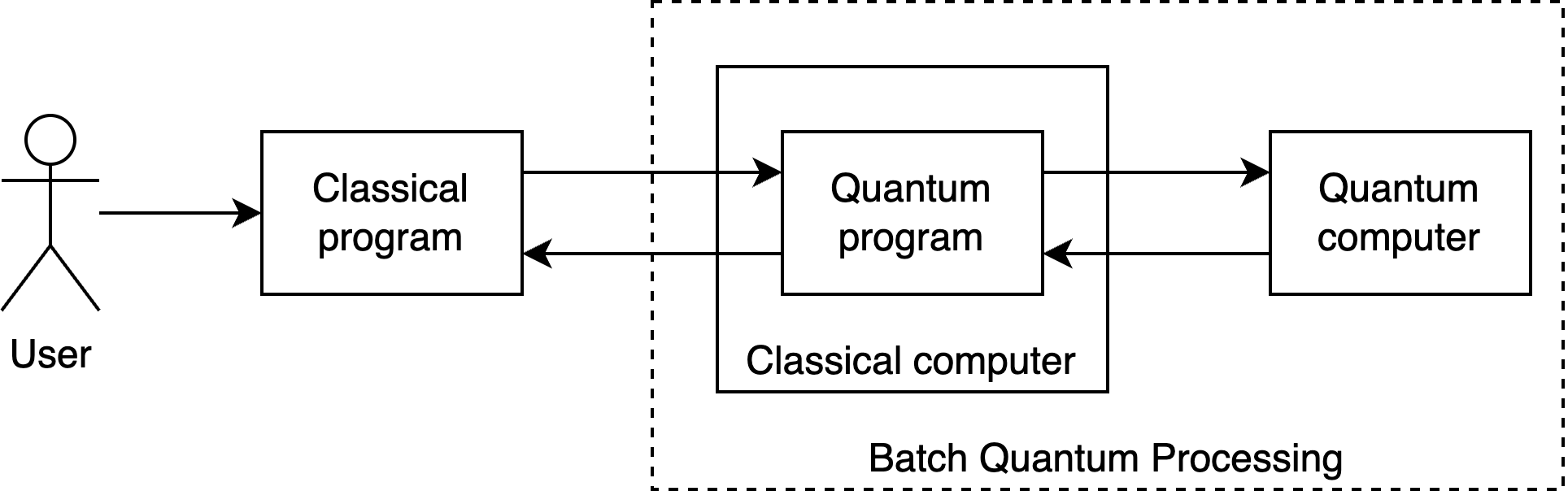}
  \caption{Quantum computing model}
  \label{fig:model}
\end{figure}

% [Low-level/circuit-level: QisKit/Cirq]
Application developer use tools like Qiskit\footnote{https://qiskit.org} and Cirq\footnote{https://quantumai.google/cirq} for writing, manipulating and optimizing quantum circuits. These Python libraries allow researchers and application developers to interact with nowadays' NISQ computers, allowing them to run quantum programs on a variety of simulators and hardware designs, abstracting away the complexities of low-level operations and allowing researchers and developers to focus on algorithm design and optimization.

% [toolkits: TensorFlow Quantum/PennyLane]
Tools like TensorFlow Quantum\footnote{https://www.tensorflow.org/quantum} and PennyLane\footnote{https://pennylane.ai} play a crucial role in facilitating the development of machine learning quantum software. These frameworks provide the high-level abstractions and interfaces that bridge the gap between quantum computing and classical machine learning. They allow researchers and developers to integrate quantum algorithms seamlessly into machine learning development process by providing access to quantum simulators and hardware, as well as offering a range of quantum-friendly classical optimization techniques. TensorFlow Quantum leverages the power of Google's TensorFlow ecosystem, enabling the combination of classical and quantum neural networks for hybrid quantum-classical machine learning models. PennyLane offers a unified framework for developing quantum machine learning algorithms, supporting various quantum devices and seamlessly integrating them with classical machine learning libraries. These tools provide a foundation for researchers to explore and experiment with quantum machine learning, accelerating the progress and adoption of quantum computing in the field of machine learning.

%[tools: Jupyter notebooks, simulators]
Jupyter Notebooks and quantum simulators play a vital role in supporting developers of quantum programs. Jupyter provides an interactive and collaborative environment where developers can write, execute, and visualize their quantum code in an accessible manner. They allow for the combination of code, explanatory text, and visualizations, making it easier to experiment, iterate, and document the development process. Quantum simulators, on the other hand, enable developers to simulate the behavior of quantum systems without the need for physical quantum hardware. These simulators provide a valuable testing ground for verifying and debugging quantum algorithms, allowing developers to gain insights into their performance and behavior before running them on actual quantum devices. Developers can iterate quickly, gain a deeper understanding of quantum concepts, and refine their quantum programs efficiently.

Traditional cloud computing providers, such as AWS Bracket\footnote{https://aws.amazon.com/braket/}, Azure Quantum\footnote{https://learn.microsoft.com/en-us/azure/quantum/}, Google Quantum AI\footnote{https://quantumai.google} or IBM Quantum\footnote{https://quantum-computing.ibm.com}, offer comprehensive quantum development services. These services are designed to optimize the development process, with integrated tools like Jupyter notebooks and task schedulers. Developers can create quantum applications and algorithms across multiple hardware platforms simultaneously. This approach ensures flexibility, allowing fine-tune algorithms for specific systems while maintaining the ability to develop applications that are compatible with various quantum hardware platforms.

\subsection{Operations}

% [SDLC]
The software development lifecycle (SDLC) of quantum programs involves a series of stages tailored to the unique challenges of quantum computing \cite{sdlc}. It typically begins with requirements gathering and problem formulation, where developers identify the specific problem that the quantum program aims to solve. During algorithm design, the developers design quantum algorithms that leverage the unique capabilities of quantum systems. The designed algorithm implementation translates the algorithm into quantum code using quantum programming languages and frameworks like Qiskit or Cirq. After implementation, the program undergoes rigorous testing and debugging, using quantum simulators to validate its functionality and behavior. The tested program is executed on actual quantum hardware, with careful consideration given to the limitations and noise inherent in quantum systems. Finally, ongoing maintenance and optimization are crucial, as quantum hardware, software frameworks, and algorithms evolve rapidly.

% [simulate HW noise, virtualization]
Simulators and virtualization offer significant advantages to quantum computing from an operations perspective. Simulators provide a virtual environment for testing and debugging quantum programs without the need for physical quantum hardware. Ops teams can validate code, identify errors, and optimize performance in a controlled and reproducible manner. Simulators also allow ops teams to simulate larger-scale quantum systems than currently available in physical hardware, providing insights into the behavior and scalability of quantum programs. Additionally, virtualization techniques enable the efficient allocation and management of quantum resources, allowing multiple users to access and share quantum computing resources securely. Ops teams can provision virtualized quantum environments, manage access controls, and monitor resource utilization effectively.

\section{Full Stack Quantum Computing}

In this section we explore the \textit{full stack} quantum computing from two perspectives: development process - looking at how they are developed, and composition - looking at how quantum applications are structurally organised and the factors that need to be considered when operationalizing the execution of applications utilizing quantum computing components.

\subsection{Development process}

\begin{figure*}[!t]
    \centering
    \includegraphics[width=0.9\textwidth]{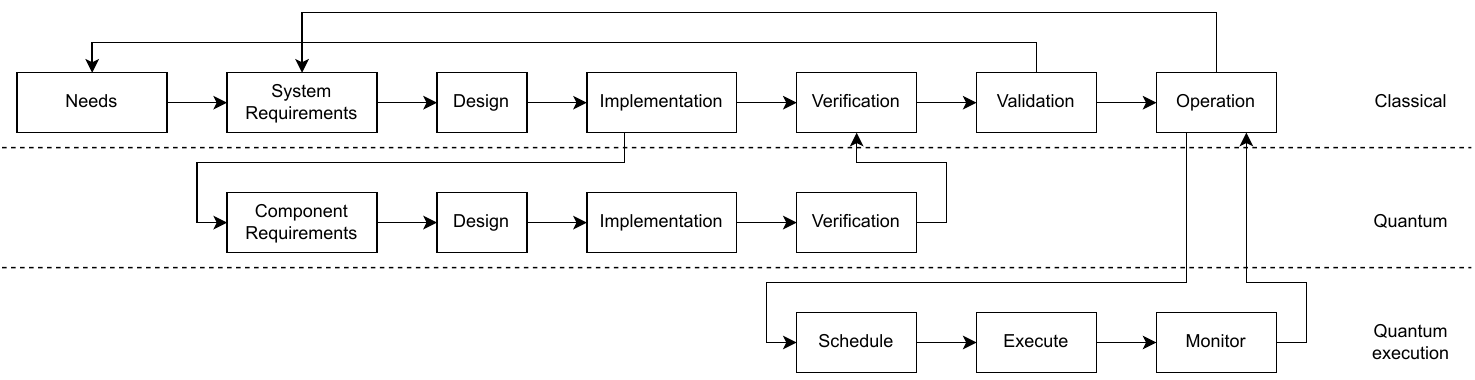}
    \caption{Software development lifecycle of a hybrid system that includes both classical and quantum technology}
    \label{fig:quantum-sdlc}
\end{figure*}

The SDLC of applications incorporating quantum technology involves streams of activities encompassing both classical and quantum components, see Fig. \ref{fig:quantum-sdlc}. At the top level, the classical software development process begins by identifying user needs and deriving system requirements. These requirements are transformed into a design and implemented, followed by verification against the requirements and validation against user needs. Once the software system enters the operational phase, any detected anomalies are used to inform potential new system requirements, if necessary. Concurrently, a dedicated track for quantum components is followed within the SDLC, specific to the implementation of quantum technology. The requirements for these components are converted into a design, which is subsequently implemented, verified, and integrated into the larger software system. The development occurs on simulators running on classical computers, which can simulate the noise characteristic of actual quantum hardware. During the operational phase, the quantum software components are executed on real hardware. Scheduling ensures efficient utilization of scarce quantum hardware, while monitoring capabilities enable the detection of anomalies throughout the process.

This workflow enables the development of products that include quantum technology using both plan-based and iterative development practices. However, when it comes to the DevOps aspects of quantum computing \cite{quantum-devops}, it becomes crucial to focus on practices and activities that facilitate effective monitoring of the quantum components operating in the production environment. %This entails implementing monitoring mechanisms to ensure the smooth functioning and performance of quantum components throughout their deployment and execution.

\subsection{Composition}

\begin{figure}
  \centering
  \includegraphics[width=0.3\textwidth]{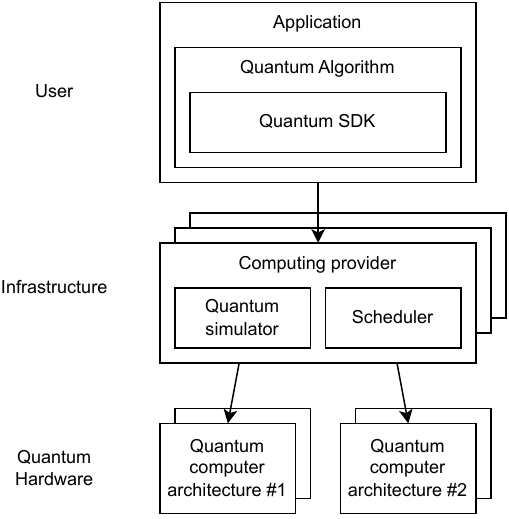}
  \caption{Ecosystem layers and relationships between stakeholders}
  \label{fig:relationships}
\end{figure}

From an architecture perspective, we can identify the following three layers: user, infrastructure and hardware (depicted in Fig. \ref{fig:relationships}). The \textit{user} software refers to the end user programs and the components developed by third parties, such as general purpose (e.g. Qiskit Terra\footnote{https://github.com/Qiskit/qiskit-terra}) or specialised (e.g. OpenFermion\footnote{https://github.com/quantumlib/OpenFermion} or TensorFlow Quantum\footnote{https://www.tensorflow.org/quantum}) libraries of quantum algorithms and circuits (e.g. Cirq and Qiskit). The \textit{infrastructure} layer contains the software needed to develop (e.g. simulators), test under realistic scenarios (e.g. simulate the noise of NISQ hardware) and run quantum programs at scale (e.g. task schedulers). The \textit{hardware} layer contains the software specific for each hardware architecture, such as the software that drives the control circuits.

\section{Goals, challenges and future research directions}

Our exploration of full-stack quantum computing focuses on identifying the challenges and difficulties in quantum software development. By leveraging the principles and practices of continuous software engineering, such as DevOps, which enable small, multidisciplinary teams to iterate quickly and deliver high-quality traditional software, we aim to pinpoint the specific components and interfaces that facilitate the transfer and application of these practices in the context of quantum software applications. Through this exercise, we seek to enhance our understanding of the pain points and opportunities for improvement in quantum software development, ultimately striving to foster the seamless integration of best practices from traditional software engineering into the emerging field of quantum computing \cite{ml-technical-depth}.

%\textbf{Challenges}: 
The main challenges emerge from two areas: technical -- integrating classical and quantum components, and process -- aligning the technical solution with user needs and requirements. These observations highlight the need to address technical and process-related hurdles in order to successfully utilize quantum technology while effectively meeting user expectations. From a development perspective, the quantum software debugging is fundamentally different than for classical software. The black box nature of the quantum computer, with its limited observability, limits the debugging capabilities. Although new quantum debugging techniques are developed \cite{debugging}, they are far from the ability to stop the execution and inspect its state at any point in time that is typically found in classical computing. Overcoming these limitations require new development approaches that require modular software development and reliable intermediate verification.

%\textbf{Future Research Directions}: 
Multiple stakeholders contribute various software and hardware components at both the classical and quantum levels. While most stakeholders focus on specific areas like quantum algorithm or hardware development, influential entities such as Google and IBM have a significant presence and influence across the entire technology stack. They are driven by diverse economic and technological interests, which can either align or conflict with one another. Similar to the design principles behind the internet \cite{tussle}, the full-stack of quantum software must be designed to accommodate these inherent conflicts by establishing well-defined trust boundaries and open interfaces. This approach that works along the tussles among the stakeholders is crucial for fostering the development of a robust commercial environment that encourages continuous investments from both public and private entities \cite{qtm}.

% Similar tussles as for the design of internet \cite{tussle} should be considered when designing quantum software systems...

% \section{Conceptual Case-study exercise}
% \label{Case study}
% We consider a conceptual exercise of a case study that discusses how the quantum computing software development process compares to the traditional software development process.   

\section{Conclusion}

Despite the novelty and the fundamentally new approach of quantum computing, the software development shares many characteristics with classical software engineering. Making reliable quantum software requires careful design that incorporates the best practices from classical computing, while focusing the development effort on specific high value components that improve the development experience and lower the operational costs.

\section*{Acknowledgement}
This work has been supported by the Academy of Finland (project DEQSE 349945) and Business Finland (project TORQS 8582/31/2022).

\bibliographystyle{ieeetr}
\bibliography{References}
\end{document}